\documentclass{article}%
\usepackage{amssymb}
\usepackage{amsmath}
\usepackage{makeidx}
\usepackage{amsfonts}
\usepackage{graphicx}%
\begin{document}

\author{SAWA MANOFF\\\textit{Bulgarian Academy of Sciences}\\\textit{\ Institute for Nuclear Research and Nuclear Energy}\\\textit{\ Department of Theoretical Physics}\\\textit{\ Blvd. Tzarigradsko Chaussee 72}\\\textit{\ 1784 Sofia - Bulgaria}}
\title{\textbf{NAVIER-STOKES' EQUATIONS FOR RADIAL AND TANGENTIAL ACCELERATIONS}}
\date{\textit{E-mail address: smanov@inrne.bas.bg}}
\maketitle

\begin{abstract}
The Navier-Stokes equations are considered by the use of the method of
Lagrangians with covariant derivatives (MLCD) over spaces with affine
connections and metrics. It is shown that the Euler-Lagrange equations appear
as sufficient conditions for the existence of solutions of the Navier-Stokes
equations over (pseudo) Euclidean and (pseudo) Riemannian spaces without
torsion. By means of the corresponding $(n-1)+1$ projective formalism the
Navier-Stokes equations for radial and tangential accelerations are found.

\end{abstract}

\section{Introduction}

By the use of the method of Lagrangians with covariant derivatives (MLCD)
\cite{Manoff-1}, \cite{Manoff-1a} the different energy-momentum tensors and
the covariant Noether's identities for a field theory as well as for a theory
of continuous media can be found. On the basis of the $(n-1)+1$ projective
formalism and by the use of the notion of covariant divergency of a tensor of
second rank the corresponding covariant divergencies of the energy-momentum
tensors could be found. They lead to Navier-Stokes' identity and to the
corresponding generalized Navier-Stokes' equations.

The general scheme for obtaining the Navier-Stokes equations for radial and
tangential acceleration could be given in the form

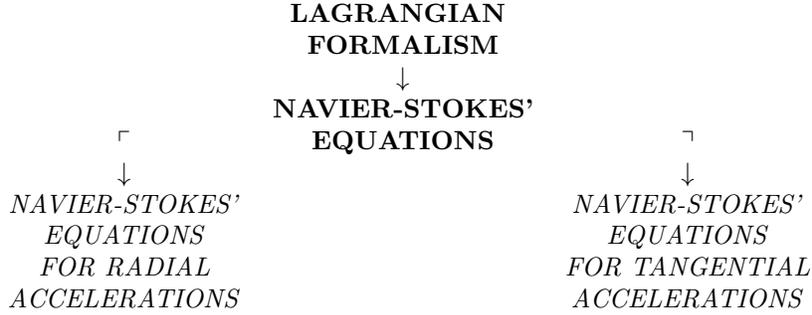
\begin{center}%
\begin{tabular}
[c]{ccc}
& \textbf{LAGRANGIAN } & \\
& \textbf{FORMALISM} & \\
& $\downarrow$ & \\
& \textbf{NAVIER-STOKES'} & \\
$\ulcorner$ & \textbf{EQUATIONS} & $\urcorner$\\
$\downarrow$ &  & $\downarrow$\\
\textit{NAVIER-STOKES'} &  & \textit{NAVIER-STOKES'}\\
\textit{EQUATIONS} &  & \textit{EQUATIONS}\\
\textit{FOR RADIAL} &  & \textit{FOR TANGENTIAL}\\
\textit{ACCELERATIONS} &  & \textit{ACCELERATIONS}%
\end{tabular}

\end{center}

The structure of a Lagrangian theory of tensor fields over a differentiable
manifold $M$ $(\dim M=n)$ could be represented in the form

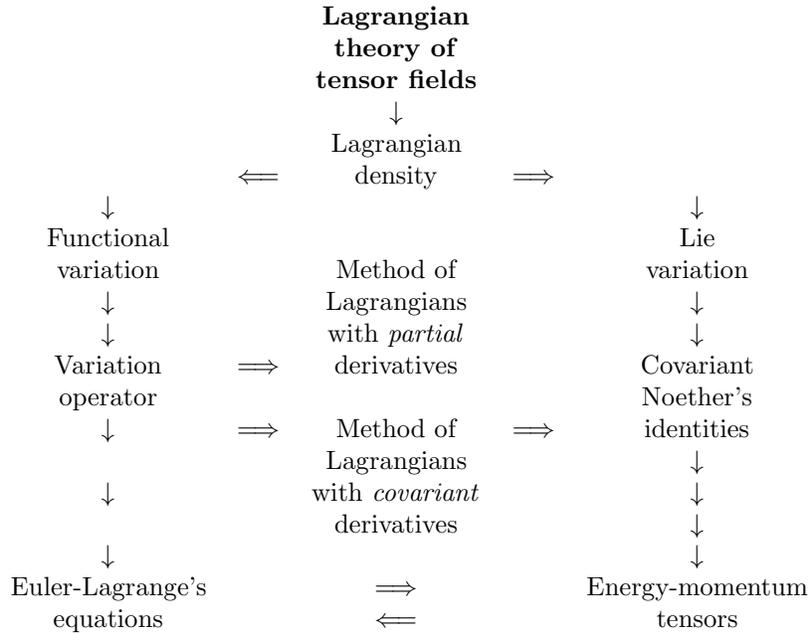
\begin{center}%
\begin{tabular}
[c]{ccccc}
&  & \textbf{Lagrangian} &  & \\
&  & \textbf{theory of} &  & \\
&  & \textbf{tensor fields} &  & \\
&  & $\downarrow$ &  & \\
&  & Lagrangian &  & \\
& $\Longleftarrow$ & density & $\Longrightarrow$ & \\
$\downarrow$ &  &  &  & $\downarrow$\\
Functional &  &  &  & Lie\\
variation &  & Method of &  & variation\\
$\downarrow$ &  & Lagrangians &  & $\downarrow$\\
$\downarrow$ &  & with \textit{partial} &  & $\downarrow$\\
Variation & $\Longrightarrow$ & derivatives &  & Covariant\\
operator &  &  &  & Noether's\\
$\downarrow$ & $\Longrightarrow$ & Method of & $\Longrightarrow$ &
identities\\
&  & Lagrangians &  & $\downarrow$\\
$\downarrow$ &  & with \textit{covariant} &  & $\downarrow$\\
&  & derivatives &  & $\downarrow$\\
$\downarrow$ &  &  &  & $\downarrow$\\
Euler-Lagrange's &  & $\implies$ &  & Energy-momentum\\
equations &  & $\Longleftarrow$ &  & tensors
\end{tabular}

\end{center}

Let the following structure
\[
(M,~V,~g,~\Gamma,~P)
\]
be given, where

(i) $M$ is a differantiable manifold with $\dim M=n$,

(ii) $V=V^{A}~_{B}\cdot e_{A}\otimes e^{B}\in\otimes^{k}~_{l}(M)$ are tensor
fields with contravariant rank $k$ and covariant rank $l$ over $M$, $A $ and
$B$ are collective indices,

(iii) $g\in\otimes_{Sym2}(M)$ is a covariant symmetric metric tensor field
over $M$,

(iv) $\Gamma$ is a contravariant affine connection, $P$ is a covariant affine
connection related to the covariant differential operator along a basis vector
field $\partial_{i}$ or $e_{i}$ in a co-ordinate or non-co-ordinate basis
respectively%
\begin{align*}
\nabla_{\partial_{i}}V  &  =V^{A}~_{B;i}\cdot\partial_{A}\otimes
dx^{B}~\text{\ \ \ \ \ ,}\\
V^{A}~_{B;i}  &  =(V^{A}~_{B,i}+\Gamma_{\operatorname{Ci}}^{A}\cdot V^{C}%
~_{B}+P_{Bi}^{D}\cdot V^{A}~_{D}\text{ ,}\\
V^{A}~_{B,i}  &  =\frac{\partial V^{A}~_{B}}{\partial x^{i}}\text{ \ \ .}%
\end{align*}

A Lagrangian density $\mathbf{L}$ can be considered in two different ways as a
tensor density of rank 0 with the weight $q=1/2$, depending on tensor field's
components and their first and second covariant derivatives

(i) As a tensor density $\mathbf{L}$ of type $1$, depending on tensor field's
components, their first (and second) \textit{partial} derivatives, (and the
components of contravariant and covariant affine connections), i.e.%
\[
\mathbf{L}=\sqrt{-d_{g}}\cdot L(g_{ij},~g_{ij,k},g_{ij,k,l},~V^{A}~_{B}%
,~V^{A}~_{B,i},~V^{A}~_{B,i,j},~\Gamma_{jk}^{i},...,~P_{jk}^{i},...)\text{
\ \ ,}%
\]
where~$L$ is a Lagrangian invariant,%
\begin{align*}
d_{g}  &  =\det(g_{ij})<0\text{ , \ \ \ \ \ }g=g_{ij}\cdot dx^{i}.dx^{j}\text{
\ , \ }\\
\text{\ }dx^{i}.dx^{j}  &  =\frac{1}{2}\cdot(dx^{i}\otimes dx^{j}%
+dx^{j}\otimes dx^{i})\text{ \ \ ,}\\
V^{A}~_{B,i,j}  &  =\frac{\partial V^{A}~_{B}}{\partial x^{j}\partial x^{i}%
}\text{ \ \ .}%
\end{align*}

The method using a Lagrangian density of type $1$ is called Method of
Lagrangians with \textit{partial} derivatives (MLPD).

(ii) As a tensor density \ $\mathbf{L}$ of type $2$, depending on tensor
field's components and their first (and second) \textit{covariant}
derivatives, i.e.%
\[
\mathbf{L}=\sqrt{-d_{g}}\cdot L(g_{ij},~g_{ij;k},g_{ij;k;l},~V^{A}~_{B}%
,~V^{A}~_{B;i},~V^{A}~_{B;i;j})\text{ \ \ .}%
\]

By the use of the variation operator $\delta$, commuting with the covariant
differential operator%
\[
\delta\circ\nabla_{\xi}=\nabla_{\xi}\circ\delta+\nabla_{\delta\xi}\text{
\ \ \ ,~\ \ \ \ }\xi\in T(M)\text{ \ \ ,~\ \ \ \ }T(M)=\cup_{x\in M}%
T_{x}(M)\text{ \ ,}%
\]
we could find the Euler-Lagrange equations.

By the use of the Lie variation operator (identical with the Lie differential
operator) $\pounds _{\xi}$, we could find the corresponding energy-momentum tensors.

The method using a Lagrangian density of type $2$ is called Method of
Lagrangians with \textit{covariant} derivatives (MLCD).

\subsection{Euler-Lagrange's equations}

The Euler-Lagrange equations follow from the variation of the Lagrangian
density of type $2$ in the form \cite{Manoff-2}

(i) for the tensor fields $V$%
\[
\frac{\delta_{v}L}{\delta V^{A}~_{B}}+P^{A}~_{B}=0\text{ \ \ ,}%
\]

(ii) for the metric tensor field $g$%
\[
\frac{\delta_{g}L}{\delta g_{kl}}+\frac{1}{2}\cdot L\cdot g^{\overline
{k}\overline{l}}+P^{kl}=0\text{ \ \ \ .}%
\]

\textit{Special cases:} (Pseudo) Euclidean and (pseudo) Riemannian spaces
without torsion.%
\[
\frac{\delta_{v}L}{\delta V^{A}~_{B}}=0\text{ \ \ , \ \ \ \ \ \ \ \ \ \ }%
\frac{\delta_{g}L}{\delta g_{kl}}+\frac{1}{2}\cdot L\cdot g^{\overline
{k}\overline{l}}=0\text{ \ \ \ .\ }%
\]

\subsection{Energy-momentum tensors}

By the use of the Lie variation operator the energy-momentum tensors follow
\cite{Manoff-2}:

(i) Generalized canonical energy-momentum tensor $\theta=\overline{\theta}%
_{i}~^{j}\cdot\partial_{j}\otimes dx^{i}$,

(ii) Symmetric energy-momentum tensor of Belinfante $_{s}T=T_{i}~^{j}%
\cdot\partial_{j}\otimes dx^{i}$,

(iii) Variational energy-momentum tensor of Euler-Lagrange $Q=\overline{Q}%
_{i}~^{j}\cdot\partial_{j}\otimes dx^{i}$.

The energy-momentum tensors obey the covariant Noether identities%
\begin{align*}
\overline{F}_{i}+\overline{\theta}_{i}~^{j}{}_{;j}  &  \equiv0~\ \ \text{,}\\
F+\delta\theta &  \equiv0\text{ \ \ \ ,}%
\end{align*}

\begin{center}
(first covariant Noether's identity)%
\begin{align*}
\overline{\theta}_{i}~^{j}-~_{s}T_{i}~^{j}  &  \equiv\overline{Q}_{i}%
~^{j}\text{ \ \ ,}\\
\theta-~_{s}T  &  \equiv Q~\ \ \ \ \text{.}%
\end{align*}

(second covariant Noether's identity)
\end{center}

Now we can draw a rough scheme of the main structure of a Lagrangian theory:

\begin{center}%
\begin{tabular}
[c]{ccccc}%
$\ulcorner$ & $\longleftarrow$ & $L$ & $\longrightarrow$ & $\urcorner$\\
$\downarrow$ &  & $\downarrow$ &  & $\downarrow$\\
$\downarrow$ &  & $_{s}T$ &  & $\delta L/\delta V^{A}~_{B}$\\
$\downarrow$ &  & $\downarrow$ &  & $\downarrow$\\
$\theta$ &  & $\downarrow$ &  & $Q$\\
$\downarrow$ &  & $\downarrow$ &  & $\downarrow$\\
$\llcorner$ & $\longrightarrow$ & $\theta-~_{s}T\equiv Q$ & $\longleftarrow$ &
$\lrcorner$\\
$\downarrow$ &  &  &  & $\downarrow$\\
$\llcorner$ & $\longrightarrow$ & $F+\delta\theta\equiv0$ & $\longleftarrow$ &
$\lrcorner$\\
&  &  &  &
\end{tabular}

\end{center}

\section{Invariant projections of the energy-momentum tensors}

By the use of the $(n-1)+1$ projective formalism we can find the invariant
projections of the energy-momentum tensors corresponding to a Lagrangian field
theory or to a theory of continuous media. The idea of the projective
formalism is the representation of the dynamic characteristics of a Lagrangian
system by means of their projections along the world line of an observer and
to local neighborhoods orthogonal to this world line. The tangent vector to
the world line of the observer and its local neighborhoods determine the
notion of frame of reference \cite{Manoff-4} $Fr(u,\tau,\xi_{\perp})$, where
$u$ is the tangent vector of the world line, $\tau$ is the parameter of the
world line, interpreted as the proper time of the observer $\xi_{\perp}$ is a
contravariant vector field, orthogonal to $u$. The variation of $\xi_{\perp}$
determines the relative velocity and the relative acceleration between the
points at the world line and the points in the neighborhoods lying in the sub
space orthogonal to the vector $u$.

Let the contravariant vector field $u\in T(M)$, $g(u,u):=e\neq0$, and its
corresponding projective metrics $h_{u}$ and $h^{u}$%
\begin{align*}
h_{u}  &  =g-\frac{1}{e}\cdot g(u)\otimes g(u)\text{ , \ \ \ \ \ \ \ \ \ \ }%
h^{u}=\overline{g}-\frac{1}{e}\cdot u\otimes u~\text{\ \ ,}\\
\overline{g}  &  =g^{ij}\cdot\partial_{i}.\partial_{j}\text{ \ \ ,
\ \ \ }\partial_{i}.\partial_{j}=\frac{1}{2}\cdot(\partial_{i}\otimes
\partial_{j}+\partial_{j}\otimes\partial_{i})~\ \text{,\ \ }%
\end{align*}
be given. Then the following proposition can be proved:

\textit{Proposition 1. }Every energy-momentum tensor $G\sim(\theta$, $_{s}T $,
$Q)$ could be represented in the form \cite{Manoff-2}%
\[
G=(\rho_{G}+\frac{1}{e}\cdot L\cdot k)\cdot u\otimes g(u)-L\cdot Kr+u\otimes
g(^{k}\pi)+\,^{k}s\otimes g(u)+(^{k}S)g\text{ ,}%
\]
where%
\[
^{k}\pi=\,^{G}\overline{\pi}\text{ ,\thinspace\thinspace\thinspace
\thinspace\thinspace\thinspace\thinspace}^{k}s=\,^{G}\overline{s}\text{
,\thinspace\thinspace\thinspace\thinspace\thinspace\thinspace\thinspace
\thinspace}^{k}S=\,^{G}\overline{S}\text{ ,}%
\]

$\rho_{G}$ is the rest mass density, $k=(1/e)\cdot\lbrack g(u)](Kr)u$, $L$ is
the pressure of the system, $Kr=g_{j}^{i}\cdot\partial_{i}\otimes dx^{j} $ is
the Kronecker tensor, $^{k}\pi$ is the conductive momentum density, $^{k}s$ is
the conductive energy flux density, ~$^{k}S$ is the stress tensor
\cite{Manoff-3}.

\section{Covariant divergency of the energy-momentum tensors and the rest mass
density}

The covariant divergency $\delta G$ of the energy-momentum tensor $\delta G$
($G\sim\,\theta$, $_{s}T$, $Q$) can be represented by the use of the
projective metrics $h^{u}$, $h_{u}$ of the contravariant vector field $u$ and
the rest mass density for the corresponding energy-momentum tensor $\rho_{G}$.

\noindent$\delta G$ and $\overline{g}(\delta G)$ can be found in the forms
\cite{Manoff-2}
\begin{equation}%
\begin{array}
[c]{c}%
\delta G=(\rho_{G}+\frac{1}{e}\cdot L\cdot k)\cdot g(a)+\\
+[u(\rho_{G}+\frac{1}{e}\cdot L\cdot k)+\,\,\,(\rho_{G}+\frac{1}{e}\cdot
L\cdot k)\cdot\delta u+\delta^{G}\overline{s}]\cdot g(u)-\\
-\,\,KrL-L\cdot\delta Kr+\delta u\cdot g(^{G}\overline{\pi})+g(\nabla
_{u}\,^{G}\overline{\pi})+g(\nabla_{^{G}\overline{s}}u)+\\
+\,\,(\rho_{G}+\frac{1}{e}\cdot L\cdot k)\cdot(\nabla_{u}g)(u)+(\nabla
_{u}g)(^{G}\overline{\pi})+(\nabla_{^{G}\overline{s}}g)(u)+\\
+\,\,\delta((^{G}\overline{S})g)\text{ , \ \ \ \ \ \ \ \ \ \ \ \ }a=\nabla
_{u}u\text{ \ \ ,}%
\end{array}
\label{X.3.-8}%
\end{equation}%
\begin{equation}%
\begin{array}
[c]{c}%
\overline{g}(\delta G)=(\rho_{G}+\frac{1}{e}\cdot L\cdot k)\cdot a+\\
+[u(\rho_{G}+\frac{1}{e}\cdot L\cdot k)+\,\,\,\,\,\,(\rho_{G}+\frac{1}{e}\cdot
L\cdot k)\cdot\delta u+\delta^{G}\overline{s}]\cdot u-\\
-\,\overline{g}(\,KrL)-L\cdot\overline{g}(\delta Kr)+\delta u\cdot
\,^{G}\overline{\pi}+\nabla_{u}\,^{G}\overline{\pi}+\nabla_{^{G}\overline{s}%
}u+\\
+\,\,(\rho_{G}+\frac{1}{e}\cdot L\cdot k)\cdot\overline{g}(\nabla
_{u}g)(u)+\overline{g}(\nabla_{u}g)(^{G}\overline{\pi})+\overline{g}%
(\nabla_{^{G}\overline{s}}g)(u)+\\
+\,\,\overline{g}(\delta((^{G}\overline{S})g))\text{ .}%
\end{array}
\label{Ch 24 ab X.3.-9}%
\end{equation}

In a co-ordinate basis $\delta G$ and $\overline{g}(\delta G)$ will have the
forms
\begin{equation}%
\begin{array}
[c]{c}%
G_{i}\,^{j}\,_{;j}=(\rho_{G}+\frac{1}{e}\cdot L\cdot k)\cdot a_{i}+\\
+\,\,[(\rho_{G}+\frac{1}{e}\cdot L\cdot k)_{,j}\cdot u^{j}+(\rho_{G}+\frac
{1}{e}\cdot L\cdot k)\cdot u^{j}\,_{;j}+\,^{G}\overline{s}^{j}\,_{;j}]\cdot
u_{i}-\\
-L_{,i}-L\cdot g_{i\,;j}^{j}+u^{j}\,_{;j}\cdot\,^{G}\overline{\pi}%
_{i}+g_{i\overline{j}}\cdot(^{G}\overline{\pi}^{j}\,_{;k}\cdot u^{k}%
+u^{j}\,_{;k}\cdot\,^{G}\overline{s}^{k})+\\
+\,\,\,g_{ij;k}\cdot\lbrack(\rho_{G}+\frac{1}{e}\cdot L\cdot k)\cdot
u^{\overline{j}}\cdot u^{k}+\,^{G}\overline{\pi}^{\overline{j}}\cdot
u^{k}+u^{\overline{j}}\cdot\,^{G}\overline{s}^{k}]+\\
+\,(g_{i\overline{k}}\cdot\,^{G}\overline{S}\,^{jk})_{;j}\text{ ,}%
\end{array}
\label{Ch 24ab X.3.-12}%
\end{equation}%
\begin{equation}%
\begin{array}
[c]{c}%
\overline{g}^{i\overline{k}}\cdot G_{k}\,^{j}\,_{;j}=(\rho_{G}+\frac{1}%
{e}\cdot L\cdot k)\cdot a^{i}+\\
+\,\,[(\rho_{G}+\frac{1}{e}\cdot L\cdot k)_{,j}\cdot u^{j}+(\rho_{G}+\frac
{1}{e}\cdot L\cdot k)\cdot u^{j}\,_{;j}+\,^{G}\overline{s}^{j}\,_{;j}]\cdot
u^{i}-\\
-L_{,j}\cdot g^{i\overline{j}}-L\cdot g^{i\overline{k}}\cdot g_{k\,;j}%
^{j}+u^{j}\,_{;j}\cdot\,^{G}\overline{\pi}^{i}+\,^{G}\overline{\pi}^{i}%
\,_{;j}\cdot u^{j}+u^{i}\,_{;j}\cdot\,^{G}\overline{s}^{j}+\\
+\,\,\,g^{i\overline{l}}\cdot g_{lj;k}\cdot\lbrack(\rho_{G}+\frac{1}{e}\cdot
L\cdot k)\cdot u^{\overline{j}}\cdot u^{k}+\,^{G}\overline{\pi}^{\overline{j}%
}\cdot u^{k}+u^{\overline{j}}\cdot\,^{G}\overline{s}^{k}]+\\
+\,g^{i\overline{l}}\cdot(g_{l\overline{k}}\cdot\,^{G}\overline{S}%
\,^{jk})_{;j}\text{ .}%
\end{array}
\label{Ch 24ab X.3.-13}%
\end{equation}

\section{Navier-Stokes' identities and Navier-Stokes' equations}

If we consider the projections of the first Noether identity along a non-null
(non-isotropic) vector field $u$ and its corresponding contravariant and
covariant projective metrics $h^{u}$ and $h_{u}$ we will find the first and
second Navier-Stokes identities.

From the Noether identities in the form
\begin{align*}
\overline{g}(F)+\overline{g}(\delta\theta)  &  \equiv0\text{,\thinspace
\thinspace\thinspace\thinspace\thinspace\thinspace\thinspace\thinspace
\thinspace\thinspace\thinspace\thinspace\thinspace\thinspace\thinspace
\thinspace\thinspace}\,\,\,\,\text{(first covariant Noether's identity)}%
\,\text{,}\,\,\\
\,\,\,\,\,\,\,\,\,\,\,\,\,\,\,\,(\theta)\overline{g}-(\,_{s}T)\overline{g}  &
\equiv(Q)\overline{g}\text{,\thinspace\thinspace\thinspace\thinspace
\thinspace\thinspace\thinspace\thinspace\thinspace\thinspace\thinspace(second
covariant Noether's identity) ,}%
\end{align*}

\noindent we can find the projections of the first Noether identity along a
contravariant non-null vector field $u=u^{i}\cdot\partial_{i}$ and orthogonal
to $u$.

Since
\begin{align}
g(\overline{g}(F),u)  &  =g_{ik}\cdot g^{\overline{k}\overline{l}}\cdot
F_{l}\cdot u^{\overline{i}}=g_{i}^{l}\cdot F_{l}\cdot u^{\overline{i}}%
=F_{i}\cdot u^{\overline{i}}=F(u)\,\text{,\thinspace\thinspace}%
\label{Ch 24ab 14.2a}\\
g(\overline{g}(\delta\theta),u)  &  =(\delta\theta)(u)\,\,\,\,\,\,\,\,\text{,
\thinspace\thinspace\thinspace\thinspace\thinspace}F=F_{k}\cdot dx^{k}\text{,}
\label{Ch 24ab 14.2b}%
\end{align}

\noindent we obtain the\textit{\ first Navier-Stokes identity} in the form
\begin{equation}
F(u)+(\delta\theta)(u)\equiv0\text{ \thinspace\thinspace.}
\label{Ch 24ab 14.3}%
\end{equation}

By the use of the relation
\begin{align}
\overline{g}[h_{u}(\overline{g})(F)]  &  =\overline{g}(h_{u}[\overline
{g}(F)])=h^{u}(F)\,\text{\thinspace\thinspace\thinspace\thinspace
\thinspace,\thinspace\thinspace\thinspace\thinspace\thinspace\thinspace
\thinspace\thinspace\thinspace\thinspace}\overline{g}(h_{u})\overline{g}%
=h^{u}\,\,\text{,}\label{Ch 24ab 14.4a}\\
\overline{g}[h_{u}(\overline{g})(\delta\theta)]  &  =\overline{g}%
(h_{u}[\overline{g}(\delta\theta)])=h^{u}(\delta\theta)\text{ ,}
\label{Ch 24ab 14.4b}%
\end{align}

\noindent the first Noether identity could be written in the forms
\begin{align}
h_{u}[\overline{g}(F)]+h_{u}[\overline{g}(\delta\theta)]  &  \equiv0\text{
\thinspace\thinspace\thinspace,}\label{Ch 24ab 14.5a}\\
h^{u}(F)+h^{u}(\delta\theta)  &  \equiv0\,\,\,\,\,\text{.}
\label{Ch 24ab 14.5b}%
\end{align}

The last two forms of the first Noether identity represent the \textit{second
Navier-Stokes identity}.

If the projection $h^{u}(F)$, orthogonal to $u$, of the volume force\thinspace
$F$ is equal to zero, we obtain the\textit{\ generalized Navier-Stokes
equation} in the form
\begin{equation}
h^{u}(\delta\theta)=0\text{ \thinspace,} \label{Ch 24ab 14.6}%
\end{equation}

\noindent or in the form
\begin{equation}
h_{u}[\overline{g}(\delta\theta)]=0\text{ \thinspace\thinspace.}
\label{Ch 24ab 14.7}%
\end{equation}

Let us now find the explicit form of the first and second Navier-Stokes
identities and the explicit form of the generalized Navier-Stokes equation.
For this purpose we can use the explicit form of the covariant divergency
$\delta\theta$ of the generalized canonical energy-momentum tensor $\theta$.

(a) The first Navier-Stokes identity follows in the form
\begin{align*}
&  F(u)+(\rho_{\theta}+\frac{1}{e}\cdot L\cdot k)\cdot g(a,u)+\\
&  +e\cdot\lbrack u(\rho_{\theta}+\frac{1}{e}\cdot L\cdot k)+(\rho_{\theta
}+\frac{1}{e}\cdot L\cdot k)\cdot\delta u+\delta^{\theta}\overline{s}]-\\
&  -(KrL)(u)-L\cdot(\delta Kr)(u)+g(\nabla_{u}\,^{\theta}\overline{\pi
},u)+g(\nabla_{^{\theta}\overline{s}}u,u)+\\
&  +(\rho_{\theta}+\frac{1}{e}\cdot L\cdot k)\cdot(\nabla_{u}g)(u,u)+(\nabla
_{u}g)(^{\theta}\overline{\pi},u)+(\nabla_{^{\theta}\overline{s}}g)(u,u)+
\end{align*}%
\begin{equation}
+[\delta((^{\theta}\overline{S})g)](u)\equiv0\,\,\,\,\,\,\,\text{.}
\label{Ch 24ab 14.8}%
\end{equation}

Since
\begin{align*}
g(u,a)  &  =\pm l_{u}\cdot\frac{dl_{u}}{d\tau}-\frac{1}{2}\cdot(\nabla
_{u}g)(u,u)=\\
&  =\frac{1}{2}\cdot\lbrack\frac{d}{d\tau}(\pm l_{u}^{2})-(\nabla
_{u}g)(u,u)]\text{ \ ,}%
\end{align*}
the first Navier-Stokes identity could be interpreted as a definition for the
change of $l_{u}^{2}$ along the world line of the observer. The length of the
non-isotropic contravariant vector $u$ is interpreted as the velocity of a
signal emitted or received by the observer \cite{Manoff-5}. On this basis, the
first Navier-Stokes identity is related to the change of the velocity of
signals emitted or received by an observer moving in a continuous media or in
a fluid.

(b) The second Navier-Stokes identity can be found in the form
\[
h_{u}[\overline{g}(F)]+h_{u}[\overline{g}(\delta\theta)]\equiv
\]%
\begin{align}
&  \equiv(\rho_{\theta}+\frac{1}{e}\cdot L\cdot k)\cdot h_{u}(a)-\nonumber\\
&  -h_{u}[\overline{g}(KrL)]-L\cdot h_{u}[\overline{g}(\delta Kr)]+\delta
u\cdot h_{u}(^{\theta}\overline{\pi})+\nonumber\\
&  +h_{u}(\nabla_{u}\,^{\theta}\overline{\pi})+h_{u}(\nabla_{^{\theta
}\overline{s}}u)+\nonumber\\
&  +(\rho_{\theta}+\frac{1}{e}\cdot L\cdot k)\cdot h_{u}[\overline{g}%
(\nabla_{u}g)(u)]+h_{u}[\overline{g}(\nabla_{u}g)(^{\theta}\overline{\pi
})]+\nonumber\\
&  +h_{u}[\overline{g}(\nabla_{^{\theta}\overline{s}}g)(u)]+h_{u}[\overline
{g}(\delta((^{\theta}\overline{S})g))]+h_{u}[\overline{g}(F)]\equiv0\text{ .}
\label{Ch 24ab 14.9}%
\end{align}

(c) The generalized Navier-Stokes equation $h_{u}[\overline{g}(\delta
\theta)]=0$ follows from the second Navier-Stokes identity under the condition
$h_{u}[\overline{g}(F)]=0$ or under the condition $F=0$%
\begin{align*}
&  (\rho_{\theta}+\frac{1}{e}\cdot L\cdot k)\cdot h_{u}(a)-\\
&  -h_{u}[\overline{g}(KrL)]-L\cdot h_{u}[\overline{g}(\delta Kr)]+\delta
u\cdot h_{u}(^{\theta}\overline{\pi})+\\
&  +h_{u}(\nabla_{u}\,^{\theta}\overline{\pi})+h_{u}(\nabla_{^{\theta
}\overline{s}}u)+\\
&  +(\rho_{\theta}+\frac{1}{e}\cdot L\cdot k)\cdot h_{u}[\overline{g}%
(\nabla_{u}g)(u)]+h_{u}[\overline{g}(\nabla_{u}g)(^{\theta}\overline{\pi})]+\\
&  +h_{u}[\overline{g}(\nabla_{^{\theta}\overline{s}}g)(u)]+h_{u}[\overline
{g}(\delta((^{\theta}\overline{S})g))]
\end{align*}%
\begin{align}
&  =0\,\,\,\,\,\,\,\text{,}\label{Ch 24ab 14.10}\\
h_{u}(a)  &  =g(a)-\frac{1}{e}\cdot g(u,a)\cdot g(u)\text{ \thinspace
\thinspace.} \label{Ch 24ab 14.11}%
\end{align}

The second Navier-Stokes identity could be considered as a definition for the
density of the inner force. If the density of the inner force is equal to
zero, i.e. if $\overline{F}=\overline{g}(F)=0$, then the covariant divergency,
$\delta\overline{\theta}=\overline{g}(\delta\theta)$ of the generalized
canonical energy-momentum tensor $\theta$ is also equal to zero, i.e.
\ $\delta\overline{\theta}=\overline{g}(\delta\theta)=0$. Then the orthogonal
to the contravariant vector field $u$ projection of the second Navier-Stokes
identity lead to the equations%
\[
\overline{g}[h_{u}(\overline{F})]=0\text{ \ \ \ \ }\Leftrightarrow\overline
{g}[h_{u}(\delta\overline{\theta})]=0\text{ \ \ \ .}%
\]

The last equation is the \textit{Navier-Stokes equation in spaces with affine
connections and metrics}. Now, we can prove the following proposition:

\textit{Proposition 2. }The necessary and sufficient condition for the
existence of the Navier-Stokes equation in a space with affine connections and
metrics is the condition for the vanishing of the density of the inner force
in a dynamic system described by the use of a Lagrangian invariant $L$,
interpreted as the pressure $p$ of the system, i.e. the necessary and
sufficient condition for
\[
\overline{g}[h_{u}(\delta\overline{\theta})]=0\text{ }%
\]

is the condition%
\[
\overline{g}[h_{u}(\overline{F})]=0\text{ \ \ \ \ \ .}%
\]

The proof follows directly from the projective second Navier-Stokes identity
$\overline{g}[h_{u}(\overline{F})]+\overline{g}[h_{u}(\delta\overline{\theta
})]\equiv0$.

\textit{Special case:} $(L_{n},g)$-spaces: $S=C$, $f^{i}\,_{j}=g_{j}^{i}$,
$g(u,u)=e=$ const. $\neq0$, $k=1$.
\begin{equation}
\delta Kr=0\text{ ,} \label{Ch 24ab 14.12}%
\end{equation}

(a) First Navier-Stokes' identity%

\begin{align*}
&  F(u)+(\rho_{\theta}+\frac{1}{e}\cdot L)\cdot g(a,u)+\\
&  +e\cdot\lbrack u(\rho_{\theta}+\frac{1}{e}\cdot L)+(\rho_{\theta}+\frac
{1}{e}\cdot L)\cdot\delta u+\delta^{\theta}\overline{s}]-\\
&  -(KrL)(u)+g(\nabla_{u}\,^{\theta}\overline{\pi},u)+g(\nabla_{^{\theta
}\overline{s}}u,u)+\\
&  +(\rho_{\theta}+\frac{1}{e}\cdot L)\cdot(\nabla_{u}g)(u,u)+(\nabla
_{u}g)(^{\theta}\overline{\pi},u)+(\nabla_{^{\theta}\overline{s}}g)(u,u)+
\end{align*}%
\begin{equation}
+[\delta((^{\theta}\overline{S})g)](u)\equiv0\,\,\,\,\,\text{.}
\label{Ch 24ab 14.13}%
\end{equation}

(b) Second Navier-Stokes' identity%

\begin{align*}
&  (\rho_{\theta}+\frac1e\cdot L)\cdot h_{u}(a)-\\
&  -h_{u}[\overline{g}(KrL)]+\delta u\cdot h_{u}(^{\theta}\overline{\pi})+\\
&  +h_{u}(\nabla_{u}\,^{\theta}\overline{\pi})+h_{u}(\nabla_{^{\theta
}\overline{s}}u)+\\
&  +(\rho_{\theta}+\frac1e\cdot L)\cdot h_{u}[\overline{g}(\nabla
_{u}g)(u)]+h_{u}[\overline{g}(\nabla_{u}g)(^{\theta}\overline{\pi})]+\\
&  +h_{u}[\overline{g}(\nabla_{^{\theta}\overline{s}}g)(u)]+h_{u}[\overline
{g}(\delta((^{\theta}\overline{S})g))]
\end{align*}
\begin{equation}
+h_{u}[\overline{g}(F)]\equiv0\text{ .} \label{Ch 24ab 14.14}%
\end{equation}

(c) Generalized Navier-Stokes' equation $h_{u}[\overline{g}(\delta\theta)]=0$%

\begin{align*}
&  (\rho_{\theta}+\frac1e\cdot L)\cdot h_{u}(a)-\\
&  -h_{u}[\overline{g}(KrL)]+\delta u\cdot h_{u}(^{\theta}\overline{\pi})+\\
&  +h_{u}(\nabla_{u}\,^{\theta}\overline{\pi})+h_{u}(\nabla_{^{\theta
}\overline{s}}u)+\\
&  +(\rho_{\theta}+\frac1e\cdot L)\cdot h_{u}[\overline{g}(\nabla
_{u}g)(u)]+h_{u}[\overline{g}(\nabla_{u}g)(^{\theta}\overline{\pi})]+\\
&  +h_{u}[\overline{g}(\nabla_{^{\theta}\overline{s}}g)(u)]+h_{u}[\overline
{g}(\delta((^{\theta}\overline{S})g))]
\end{align*}
\begin{equation}
=0\,\,\,\,\,\,\,\text{.} \label{Ch 24ab 14.15}%
\end{equation}

\textit{Special case: }$V_{n}$-spaces: $S=C$, $f^{i}\,_{j}=g_{j}^{i}$,
$\nabla_{\xi}g=0 $ for $\forall\xi\in T(M)$, $g(u,u)=e=$ const. $\neq0$,
$k=1$, $g(a,u)=0$.

(a) First Navier-Stokes' identity%

\begin{align*}
&  F(u)+\\
&  +e\cdot[u(\rho_{\theta}+\frac1e\cdot L)+(\rho_{\theta}+\frac1e\cdot
L)\cdot\delta u+\delta^{\theta}\overline{s}]-(KrL)(u)+
\end{align*}
\begin{equation}
+[\delta((^{\theta}\overline{S})g)](u)\equiv0\,\,\,\,\,\text{.}
\label{Ch 24ab 14.16}%
\end{equation}

(b) Second Navier-Stokes' identity%

\begin{equation}
(\rho_{\theta}+\frac1e\cdot L)\cdot h_{u}(a)-h_{u}[\overline{g}(KrL)]+\delta
u\cdot h_{u}(^{\theta}\overline{\pi})+h_{u}[\overline{g}(\delta((^{\theta
}\overline{S})g))]+h_{u}[\overline{g}(F)]\equiv0\text{ .}
\label{Ch 24ab 14.17}%
\end{equation}

(c) Generalized Navier-Stokes' equation $h_{u}[\overline{g}(\delta\theta)]=0$%

\begin{equation}
(\rho_{\theta}+\frac1e\cdot L)\cdot h_{u}(a)-h_{u}[\overline{g}(KrL)]+\delta
u\cdot h_{u}(^{\theta}\overline{\pi})+h_{u}[\overline{g}(\delta((^{\theta
}\overline{S})g))]=0\text{ .} \label{Ch 24ab 14.18}%
\end{equation}

If we express the stress (tension) tensor $(^{\theta}\overline{S})g$ by the
use of the shear stress tensor $_{ks}\overline{D}$, rotation (vortex) stress
tensor $_{k}\overline{W}$, and the expansion stress invariant $_{k}%
\overline{U}$ then the covariant divergency of the corresponding tensors could
be found and at the end we will have the explicit form of the Navier-Stokes
identities and the generalized Navier-Stokes' equation including all necessary
tensors for further applications. The way of obtaining the Navier-Stokes
equations could be given in the following rough scheme

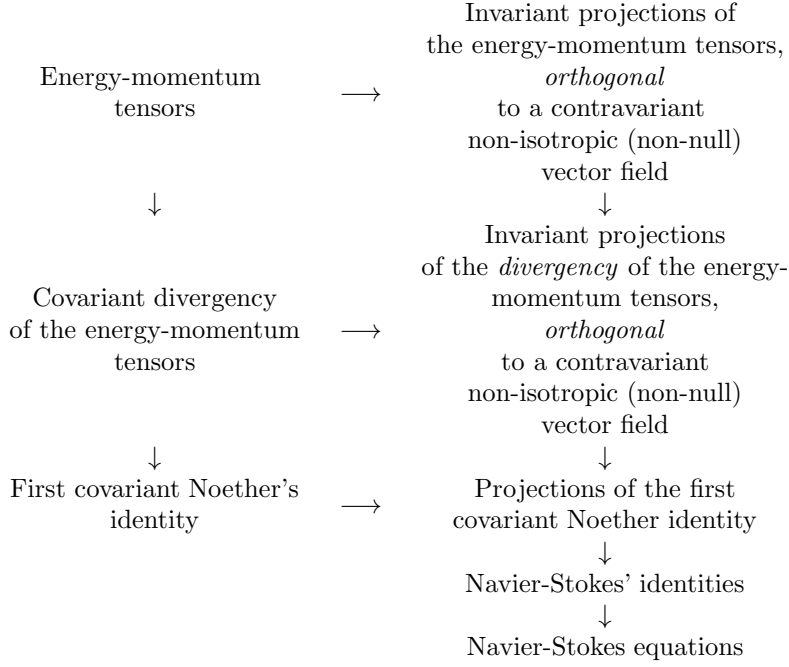
\begin{center}
$%
\begin{array}
[c]{ccc}%
\begin{array}
[c]{c}%
\text{Energy-momentum }\\
\text{tensors}%
\end{array}
& \longrightarrow &
\begin{array}
[c]{c}%
\text{Invariant projections of }\\
\text{the energy-momentum tensors,}\\
\text{\textit{orthogonal}}\\
\text{to a contravariant }\\
\text{non-isotropic (non-null)}\\
\text{vector field}%
\end{array}
\\
\downarrow &  & \downarrow\\%
\begin{array}
[c]{c}%
\text{Covariant divergency}\\
\text{of the energy-momentum}\\
\text{tensors}%
\end{array}
& \longrightarrow &
\begin{array}
[c]{c}%
\text{Invariant projections}\\
\text{of the \textit{divergency} of the energy-}\\
\text{momentum tensors,}\\
\text{\textit{orthogonal}}\\
\text{to a contravariant}\\
\text{non-isotropic (non-null)}\\
\text{vector field}%
\end{array}
\\
\downarrow &  & \downarrow\\%
\begin{array}
[c]{c}%
\text{First covariant Noether's}\\
\text{identity}%
\end{array}
& \longrightarrow &
\begin{array}
[c]{c}%
\text{Projections of the first}\\
\text{covariant Noether identity}%
\end{array}
\\
&  & \downarrow\\
&  & \text{Navier-Stokes' identities}\\
&  & \downarrow\\
&  & \text{Navier-Stokes equations}\\
&  &
\end{array}
$
\end{center}

\section{Invariant projections of Navier-Stokes' equations}

\subsection{Navier-Stokes' equations \ and Euler-Lagrange's equations}

Let us now consider the second Navier-Stokes identity in the form
\cite{Manoff-2}
\[
\overline{g}[h_{u}[\overline{g}(F)]]+\overline{g}[h_{u}[\overline{g}%
(\delta\theta)]]\equiv0
\]
or in the form%
\begin{align*}
F_{\perp}+\delta\theta_{\perp}  &  \equiv0\text{ \ \ \ , \ \ \ \ }\\
\text{\ }F_{\perp}  &  =\overline{g}[h_{u}[\overline{g}(F)]]\text{ \ \ ,
\ \ \ }\delta\theta_{\perp}=\overline{g}[h_{u}[\overline{g}(\delta
\theta)]\text{ \ \ ,}\\
g(u,F_{\perp})  &  =0\text{ \ \ \ , \ \ \ \ \ \ }g(u,\delta\theta_{\perp
})=0\text{ \ .}%
\end{align*}

The explicit form of the density $F$ of the inner force could be given as
\cite{Manoff-2}%
\begin{align*}
F  &  =\overline{F}_{i}\cdot dx^{i}\text{ \ \ \ ,}\\
\overline{F}_{i}  &  =\frac{\delta L}{\delta V^{A}~_{B}}\cdot V^{A}%
~_{B;i}+W_{i}\text{ \ \ ,}\\
W_{i}  &  =W_{i}(T_{kl}~^{j}\text{, \ \ }g_{jk;l})\text{ \ ,}%
\end{align*}
where $T_{kl}~^{j}$ are the components of the torsion tensor (in a co-ordinate
basis $T_{kl}~^{j}=\Gamma_{lk}^{j}-\Gamma_{kl}^{j}$).

For (pseudo) Euclidean and (pseudo) Riemannian spaces without torsion
$(T_{kl}^{i}=0)$ the quantity $W$ \ is equal to zero $(W_{i}=0)$ and the
density of the inner force $F$ has the form%
\[
\overline{F}_{i}=\frac{\delta L}{\delta V^{A}~_{B}}\cdot V^{A}~_{B;i}%
\]

If the Euler-Lagrange equations are fulfilled in (pseudo) Euclidean and
(pseudo) Riemannian spaces without torsion, i.e. if%
\[
\frac{\delta L}{\delta V^{A}~_{B}}=0\text{ ,}%
\]
then $F=0$ and%
\[
\overline{F}_{i}=\frac{\delta L}{\delta V^{A}~_{B}}\cdot V^{A}~_{B;i}=0\text{
\ ,}%
\]
and the following propositions could be proved:

\textit{Proposition 3. }Sufficient conditions for the existence of the
Navier-Stokes equation in (pseudo) Euclidean and (pseudo) Riemannian spaces
without torsion are the Euler-Lagrange equations.

\textit{Proposition 4}. \ Every contravariant vector field $u\in T(M)$ in
(pseudo) Euclidean and (pseudo) Riemannian spaces without torsion is a
solution of the Navier-Stokes equation if the Euler-Lagrange equations are
fulfilled for the dynamic system, described by a given Lagrangian invariant
$L=p$ interpreted as the pressure of the system.

\textit{Corollary.} If $L=p=p(u^{i}$, $u^{i}~_{;j}$, $u^{i}~_{;j;k}$, $g_{ij}
$, $g_{ij;k}$, $g_{ij;k;l}$, $V^{A}~_{B}$, $V^{A}~_{B;i}$, $V^{A}~_{B;i;j})$
is a Lagrangian density fulfilling the Euler-Lagrange equations for $u^{i}%
~$and $V^{A}~_{B}$ in (pseudo) Euclidean and (pseudo) Riemannian spaces
without torsion, then the contravariant non-isotropic vector field $u$ is also
a solution of the Navier-Stokes equation.

\subsection{Representation of $F_{\perp}$ and $\delta\theta_{\perp}$}

Now, we can use the corresponding to a vector field $\xi_{\perp}$,
$g(u,\xi_{\perp})=0$ (orthogonal to the vector field $u$) projective metrics
$h_{\xi_{\perp}}$ and $h^{\xi_{\perp}}$%
\begin{align*}
h_{\xi_{\perp}}  &  =g-\frac{1}{g(\xi_{\perp},\xi_{\perp})}\cdot g(\xi_{\perp
})\otimes g\left(  \xi_{\perp}\right)  \text{ \ \ \ ,}\\
h^{\xi_{\perp}}  &  =\overline{g}-\frac{1}{g(\xi_{\perp},\xi_{\perp})}\cdot
\xi_{\perp}\otimes\xi_{\perp}\text{ \ \ \ .}%
\end{align*}

The vector field $F_{\perp}$ could be written in the form \cite{Manoff-3},
\cite{Manoff-4}%
\begin{align*}
F_{\perp}  &  =\frac{g(F_{\perp},\xi_{\perp})}{g(\xi_{\perp},\xi_{\perp}%
)}\cdot\xi_{\perp}+\overline{g}[h_{\xi_{\perp}}(F_{\perp})]=\mp g(F_{\perp
},n_{\perp})\cdot n_{\perp}+\overline{g}[h_{\xi_{\perp}}(F_{\perp})]=\\
&  =F_{\perp z}+F_{\perp c}\text{ \ \ , \ \ \ \ }\xi_{\perp}=l_{\xi_{\perp}%
}\cdot n_{\perp}\text{ \ , \ \ \ \ }g(n_{\perp},n_{\perp})=\mp1\text{ \ .}%
\end{align*}

$F_{\perp z}$ is the radial inner force density and $F_{\perp c}$ is the
tangential (Coriolis) inner force density%
\begin{align*}
F_{\perp z}  &  =\mp g(F_{\perp},n_{\perp})\cdot n_{\perp}\text{ \ ,
\ \ \ }F_{\perp c}=\overline{g}[h_{\xi_{\perp}}(F_{\perp})]\text{ \ ,}\\
g(F_{\perp z},u)  &  =0\text{ \ , \ \ \ \ }g(F_{\perp c},\xi_{\perp})=0\text{
\ \ , \ \ \ \ \ \ }g(F_{\perp c},u)=0\text{\ \ \ .}%
\end{align*}

The Navier-Stokes equation could now be written in the form%
\[
\delta\theta_{\perp}=\mp g(\delta\theta_{\perp},n_{\perp})\cdot n_{\perp
}+\overline{g}[h_{\xi_{\perp}}(\delta\theta_{\perp})]=0\text{ \ ,}%
\]
or in the forms%
\begin{align*}
\delta\theta_{\perp z}  &  :=\mp g(\delta\theta_{\perp},n_{\perp})\cdot
n_{\perp}=0\text{ ,}\\
&  \text{Navier-Stokes' equation for radial accelerations ,}%
\end{align*}%
\begin{align*}
\delta\theta_{\perp c}  &  :=\overline{g}[h_{\xi_{\perp}}(\delta\theta_{\perp
})]=0\text{ \ \ \ ,}\\
&  \text{Navier-Stokes' equation for tangential accelerations.}%
\end{align*}

\subsection{Radial projections of Navier-Stokes' equation. Navier-Stokes'
equation for radial accelerations}

If we use the explicit form of the Navier-Stokes equation%

\begin{align}
&  (\rho_{\theta}+\frac{1}{e}\cdot L\cdot k)\cdot a_{\perp}-\nonumber\\
&  -[\overline{g}(KrL)]_{\perp}-L\cdot\lbrack\overline{g}(\delta Kr)]_{\perp
}+\delta u\cdot~^{\theta}\overline{\pi}_{\perp}+\nonumber\\
&  +(\nabla_{u}\,^{\theta}\overline{\pi})_{\perp}+(\nabla_{^{\theta}%
\overline{s}}u)_{\perp}+\nonumber\\
&  +(\rho_{\theta}+\frac{1}{e}\cdot L\cdot k)\cdot\lbrack\overline{g}%
(\nabla_{u}g)(u)]_{\perp}+[\overline{g}(\nabla_{u}g)(^{\theta}\overline{\pi
})]_{\perp}+\nonumber\\
&  +[\overline{g}(\nabla_{^{\theta}\overline{s}}g)(u)]_{\perp}+[\overline
{g}(\delta((^{\theta}\overline{S})g))]_{\perp}=0\text{ ,}\\
L  &  =p\text{ ,}%
\end{align}
and apply the projection of the Navier-Stokes equation along and orthogonal to
the vector field $\xi_{\perp}$, by the use of the representation of the
acceleration $a_{\perp}$ in the form%
\begin{align*}
a_{\perp}  &  =g(a_{\perp},n_{\perp})\cdot n_{\perp}+\overline{g}%
[h_{\xi_{\perp}}(a_{\perp})]=a_{z}+a_{c}\text{ \ ,}\\
a_{z}  &  =g(a_{\perp},n_{\perp})\cdot n_{\perp}\text{ \ \ \ \ ,
\ \ \ \ \ \ }a_{c}=\overline{g}[h_{\xi_{\perp}}(a_{\perp})]\text{ \ ,}%
\end{align*}
where $a_{z}=g(a_{\perp},n_{\perp})\cdot n_{\perp}=\mp l_{a_{z}}\cdot
n_{\perp}$ is the radial (centrifugal, centripetal) acceleration and $a_{c}= $
$\overline{g}[h_{\xi_{\perp}}(a_{\perp})]=\mp l_{a_{c}}\cdot m_{\perp}$,
$g(n_{\perp},m_{\perp})=0$, \ is the tangential (Coriolis) acceleration, we
could fine the explicit form of the Navier-Stokes equation for radial
(centrifugal, centripetal) accelerations in the form%

\begin{align}
&  (\rho_{\theta}+\frac{1}{e}\cdot L\cdot k)\cdot a_{z}-\nonumber\\
&  -[\overline{g}(KrL)]_{\perp z}-L\cdot\lbrack\overline{g}(\delta Kr)]_{\perp
z}+\delta u\cdot~^{\theta}\overline{\pi}_{\perp z}+\nonumber\\
&  +(\nabla_{u}\,^{\theta}\overline{\pi})_{\perp z}+(\nabla_{^{\theta
}\overline{s}}u)_{\perp z}+\nonumber\\
&  +(\rho_{\theta}+\frac{1}{e}\cdot L\cdot k)\cdot\lbrack\overline{g}%
(\nabla_{u}g)(u)]_{\perp z}+[\overline{g}(\nabla_{u}g)(^{\theta}\overline{\pi
})]_{\perp z}+\nonumber\\
&  +[\overline{g}(\nabla_{^{\theta}\overline{s}}g)(u)]_{\perp z}+[\overline
{g}(\delta((^{\theta}\overline{S})g))]_{\perp z}=0\text{ ,}\\
L  &  =p\text{ ,}%
\end{align}

\textit{Special case:} Perfect fluids: $^{\theta}\overline{\pi}=0$, $^{\theta
}\overline{s}=0$, $^{\theta}\overline{S}=0$, $L=p$.%

\begin{align*}
&  (\rho_{\theta}+\frac{1}{e}\cdot L\cdot k)\cdot a_{z}-[\overline
{g}(KrL)]_{\perp z}-L\cdot\lbrack\overline{g}(\delta Kr)]_{\perp z}+\\
&  +(\rho_{\theta}+\frac{1}{e}\cdot L\cdot k)\cdot\lbrack\overline{g}%
(\nabla_{u}g)(u)]_{\perp z}%
\end{align*}

\begin{equation}
=0\text{ .}%
\end{equation}

\textit{Special case: }Perfect fluids in (pseudo) Euclidean and (pseudo)
Riemannian spaces without torsion: $^{\theta}\overline{\pi}=0$, $^{\theta
}\overline{s}=0$, $^{\theta}\overline{S}=0$, $L=p$, $\nabla_{u}g=0$, $\delta
Kr=0$.%
\[
(\rho_{\theta}+\frac{1}{e}\cdot p)\cdot a_{z}=[\overline{g}(Krp)]_{\perp
z}\text{ \ ,}%
\]%
\[
a_{z}=\frac{1}{\rho_{\theta}+\frac{1}{e}\cdot L}\cdot\lbrack\overline
{g}(Krp)]_{\perp z}\text{ \ \ .}%
\]

\subsection{Tangential projections of Navier-Stokes' equation. Navier-Stokes'
equation for tangential accelerations}

For tangential (Coriolis') accelerations the Navier-Stokes equation takes the
form \cite{Manoff-3}%

\begin{align}
&  (\rho_{\theta}+\frac{1}{e}\cdot L\cdot k)\cdot a_{c}-\nonumber\\
&  -[\overline{g}(KrL)]_{\perp c}-L\cdot\lbrack\overline{g}(\delta Kr)]_{\perp
c}+\delta u\cdot~^{\theta}\overline{\pi}_{\perp c}+\nonumber\\
&  +(\nabla_{u}\,^{\theta}\overline{\pi})_{\perp c}+(\nabla_{^{\theta
}\overline{s}}u)_{\perp c}+\nonumber\\
&  +(\rho_{\theta}+\frac{1}{e}\cdot L\cdot k)\cdot\lbrack\overline{g}%
(\nabla_{u}g)(u)]_{\perp c}+[\overline{g}(\nabla_{u}g)(^{\theta}\overline{\pi
})]_{\perp c}+\nonumber\\
&  +[\overline{g}(\nabla_{^{\theta}\overline{s}}g)(u)]_{\perp c}+[\overline
{g}(\delta((^{\theta}\overline{S})g))]_{\perp c}=0\text{ ,}\\
L  &  =p\text{ .}%
\end{align}

\textit{Special case:} Perfect fluids: $^{\theta}\overline{\pi}=0$, $^{\theta
}\overline{s}=0$, $^{\theta}\overline{S}=0$, $L=p$.%

\begin{align}
&  (\rho_{\theta}+\frac{1}{e}\cdot L\cdot k)\cdot a_{c}-\nonumber\\
&  -[\overline{g}(KrL)]_{\perp c}-L\cdot\lbrack\overline{g}(\delta Kr)]_{\perp
c}+(\rho_{\theta}+\frac{1}{e}\cdot L\cdot k)\cdot\lbrack\overline{g}%
(\nabla_{u}g)(u)]_{\perp c}=0\text{ .}%
\end{align}

\textit{Special case:} Perfect fluids in (pseudo) Euclidean and (pseudo)
Riemannian spaces without torsion: $^{\theta}\overline{\pi}=0$, $^{\theta
}\overline{s}=0$, $^{\theta}\overline{S}=0$, $L=p$, $\nabla_{u}g=0$, $\delta
Kr=0$.%

\begin{align}
&  (\rho_{\theta}+\frac{1}{e}\cdot p)\cdot a_{c}=\nonumber\\
&  =[\overline{g}(Krp)]_{\perp c}\text{ \ ,}\\
a_{c}  &  =\frac{1}{(\rho_{\theta}+\frac{1}{e}\cdot p)}\cdot\lbrack
\overline{g}(Krp)]_{\perp c}\text{ \ .}%
\end{align}

\section{Conclusions}

The representations of the Navier-Stokes equation in its forms for radial
(centrifugal, centripetal) and tangential (Coriolis') accelerations could be
used for description of different motions of fluids and continuous media in
continuous media mechanics, in hydrodynamics \ and in astrophysics. The method
of Lagrangians with covariant derivatives (MLCD) appears to be a fruitful tool
for working out the theory of continuous media mechanics and the theory of
fluids in spaces with affine connections and metrics, considered as
mathematical models of space-time.


\begin{thebibliography}{9}                                                                                                %


\bibitem {Manoff-1}Manoff S., \textit{Geometry and Mechanics in Different
Models of Space-Time}: \textit{Geometry and Kinematics}. (New York: Nova
Science Publishers, 2002) Parts 2 - 3

\bibitem {Manoff-1a}Manoff S., \textit{\ Spaces with contravariant and
covariant affine connections and metrics,} Physics of Elementary Particles and
Nuclei (Particles and Nuclei) [Russian Edition: \textbf{30} (1999)\ 5,
1211-1269], [English Edition: \textbf{30} (1999) 5, 527-549]

\bibitem {Manoff-2}Manoff S., \textit{Geometry and Mechanics in Different
Models of Space-Time}: \textit{Dynamics and Applications}. (New York: Nova
Science Publishers, 2002) Parts 1 - 2

\bibitem {Manoff-3}Manoff S., \textit{Centrifugal (centripetal), Coriolis
velocities, accelerations, and Hubble law in spaces with affine connections
and metrics,} Central European J. of Physics \textbf{4} (2003)\ 660-694
(Preprint: ArXiv gr-qc/02 12 038)

\bibitem {Manoff-4}Manoff S., \textit{Frames of reference in spaces with
affine connections and metrics}, \textit{Class. Quantum Grav}. \textbf{18}
(2001)\ 6, 1111-1125. (Preprint: ArXiv gr-qc/99 08 061)

\bibitem {Manoff-5}Manoff S., \textit{Propagation of signals in spaces with
affine connections and metrics as models of space-time}, Physics of elementary
particles and nuclei (Particles and Nuclei) \textbf{35} (2004) 5, 1185-1258
(Preprint: ArXiv gr-qc/03 09 050)
\end{thebibliography}
\end{document}